\documentclass[11pt]{article}% Please use this class with these options

\usepackage{amssymb,latexsym,amsmath,amsfonts,subfigure}
\usepackage{subfigure}
\usepackage{graphicx, epsf}
\usepackage[active]{srcltx}

\numberwithin{equation}{section}

\begin{document}

\title{Post-Double Hopf Bifurcation Dynamics and Adaptive Synchronization of a Hyperchaotic System}

\author{G. Gambino\footnote{Department of Mathematics, University of Palermo, Italy, gaetana@math.unipa.it} $\;$
and S.Roy Choudhury\footnote{Department of Mathematics, University of Central Florida, USA, choudhur@cs.ucf.edu} $\;$
}

%\titlerunning{Post-Double Hopf Bifurcation Dynamics and Adaptive Synchronization}        

\maketitle

\begin{abstract}
In this paper a four-dimensional hyperchaotic system with only
one equilibrium is considered and its double Hopf bifurcations are investigated.
The general post-bifurcation and stability analysis are carried out using the normal form of the
system obtained via the method of multiple scales. The dynamics of
the orbits predicted through the normal form comprises possible regimes of
periodic solutions, two-period tori, and three-period tori in parameter space.

Moreover, we show how the hyperchaotic synchronization of this system
can be realized via an adaptive control scheme. Numerical simulations are
included to show the effectiveness of
the designed control.

%\keywords{Hyperchaotic System \and Double-Hopf Bifurcation \and Normal Form \and Adaptive synchronization}
\end{abstract}

\section{Introduction}
\label{intro}

In the last decades study of hyperchaos has received great attention
due to its important role in nonlinear science \cite{Ross,MC,Yan}. Several hyperchaotic systems with
more than one positive Lyapunov exponent have been constructed via control
techniques.

In this paper we consider a four dimensional dynamical system
firstly introduced in \cite{ref9}. This system was constructed by
adding a linear feedback controller to the second equation of the
chaotic Chen system \cite{Chen} and an additional new state
equation. The rich dynamics of this modified driven Chen system has been observed in
\cite{GGC} via computer simulations showing both chaotic and
hyperchaotic attractors and period-doubling bifurcations.

The aim of this paper is, from one hand, to understand better the onset of the hyperchaos by
the investigation of double Hopf bifurcation of
the system and the analysis of the post-bifurcation dynamics via the normal form theory \cite{N}, \cite{NB}.
From the other hand we want to realize both the chaotic and the hyperchaotic synchronization
of the system designing a global adaptive controller.

We will use in Section 3 a perturbation technique, essentially based on multiple
scales method, to compute the post-double Hopf normal form \cite{Yu1,Yu2,Yu5}. This method does not need the application of the Center
Manifold Theory and can be employed with the aid of symbolic
computation \cite{Yu3,Yu4}.

Numerical simulations are performed to corroborate the predictions from
the normal form, revealing the existence of stable periodic and toroidal attractors in the post-supercritical-Hopf cases.

Moreover, since the interest into synchronization of hyperchaotic systems has been always increasing, in particular due to its applications in secure communications (in fact the presence of more than one Lyapunov exponent generates more complex dynamics and improves the security), we propose a scheme to synchronize the modified hyperchaotic Chen system via adaptive control.
Recently, many methods and techniques have been developed to realize synchronization and control of chaotic systems, see references \cite{ref1,ref2,ref3,GLS1,ref4,ref5} and \cite{ref7} for a review. In the same spirit of \cite{Park05}, in Section 4 we design an adaptive control law and an update rule for uncertain parameters based on Lyapunov stability theory.

Numerical simulations are presented to verify the effectiveness of the proposed synchronization method both in chaotic and hyperchaotic regime.

\section{Linear analysis and double Hopf bifurcations}
\setcounter{figure}{0}
\setcounter{equation}{0}

Let us consider the following modified Chen system, firstly obtained in \cite{ref9}:

\begin{equation}\left\{
\begin{array}{l}\label{os}
\dot{x_1}=a(x_2-x_1)\\
\dot{x_2}=-dx_1-x_1x_3+cx_2-x_4\\
\dot{x_3}=x_1x_2-bx_3\\
\dot{x_4}=x_1+k
\end{array}\right.
\end{equation}

\noindent where $a,b,c,d$ and $k$ are real constants.

Once introduced the following coordinates transformations:

$$
x=x_1+k,\ \ y=y_1+k,\ \ z=x_3-\frac{k^2}{b},\ \ w=x=x_4-k\left(d-c+\frac{k^2}{b}\right),
$$

\noindent which translate to the origin the only equilibrium $E$:

\begin{equation}
E\equiv \left(-k,
-k, \frac{k^2}{b}, k\left(d-c+\frac{k^2}{b}\right)\right),
\end{equation}

\noindent the system (\ref{os}) becomes:

\begin{equation}\left\{
\begin{array}{l}\label{os0}
\dot{x}=a(y-x)\\
\dot{y}=-dx-xz+cy-w-\displaystyle\frac{k^2}{b}x+kz\\
\dot{z}=xy-bz-k(x+y)\\
\dot{w}=x
\end{array}\right.
\end{equation}

\begin{figure}[h]
\begin{center}
\subfigure[] {\epsfxsize=1.9 in \epsfbox{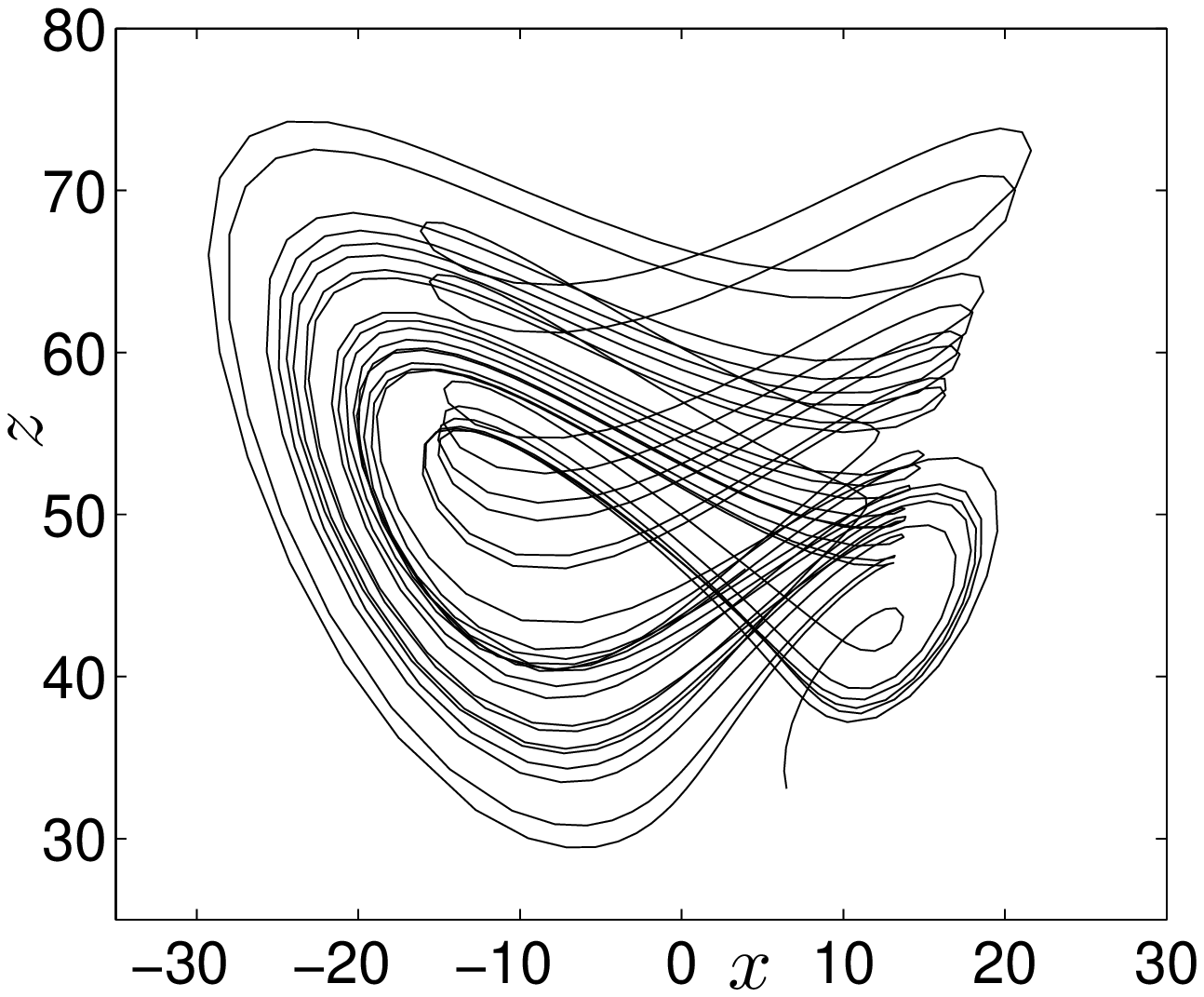}}
\subfigure[] {\epsfxsize=1.9 in \epsfbox{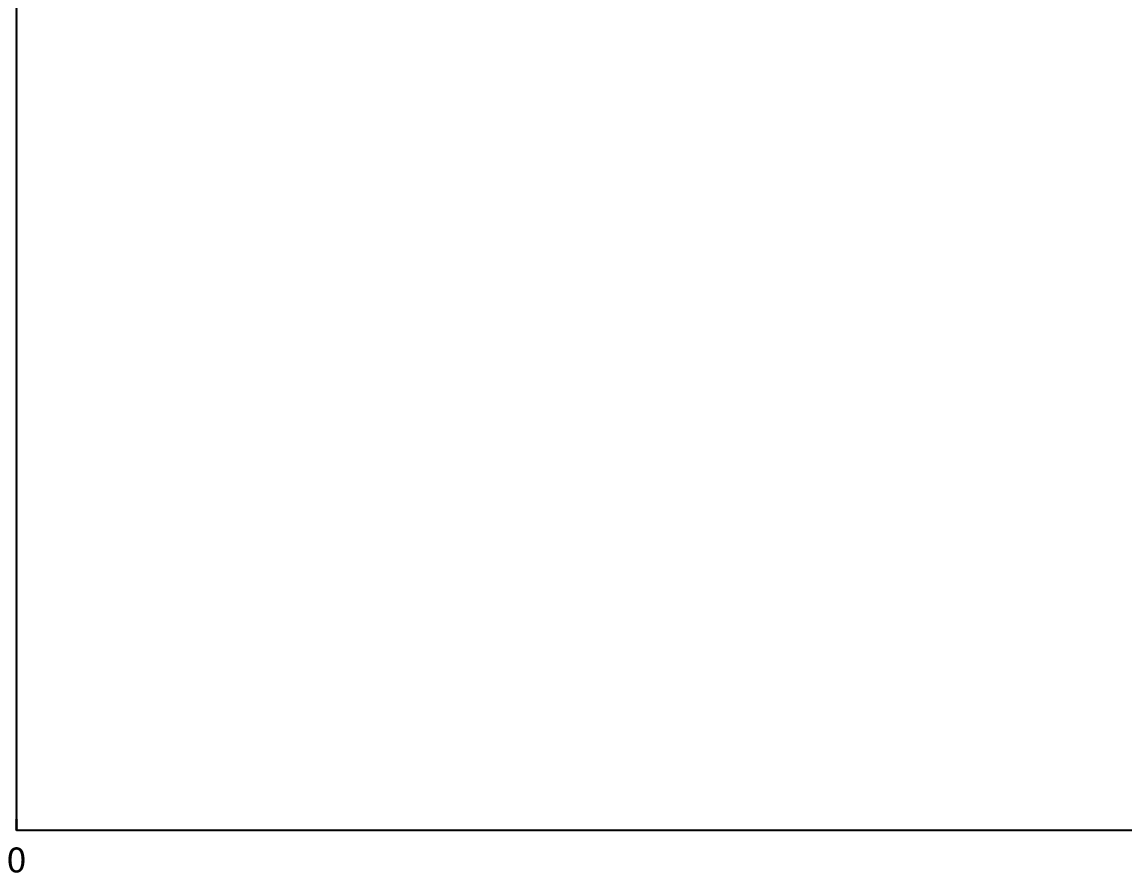}}
\end{center}
\caption{\label{f1} The parameters are chosen as
$a=36,\, b=3,\,c=28, d=-16$ and $k=1$. (a) The hyperchaotic attractor. (b) The dynamics of the Lyapunov exponents: the system admits two positive Lyapunov exponents.}
\end{figure}

\noindent By choosing the values of parameters $a=36,\, b=3,\,c=28$ and $d=-16$ the
origin is unstable and $\nabla V=-a-b+c<0$. Therefore the system is dissipative and the trajectories converge to an
attractor which is hyperchaotic, as shown in Fig. \ref{f1}.

\noindent Several bifurcation
routes to hyperchaos from periodic, quasi-periodic and chaotic
orbits are observed in \cite{GGC} via computer simulations. Since there exists only one equilibrium point, the possibility of the fixed point undergoing transcritical, pitchfork or saddle-node bifurcations (all of which involve creation, annihilation or exchange of stability of a pair of equilibria) is precluded. Thus the origin can be only destabilized via a Hopf or double Hopf bifurcation.

Applying the Routh-Hurwitz criterion one gets the following conditions for the origin being a stable equilibrium:

\begin{equation}\label{sc}
a_3>0,\ \  a_0>0,\ \ a_3a_2-a_1>0,\ \ a_3(a_2a_1-a_3a_0)-a_1^2>0,
\end{equation}

\noindent where:

\begin{eqnarray*}
a_3&=&a+b-c\\
a_2&=&ab+\frac{ak^2}{b}-ac+ad-bc+k^2\\
a_1&=&a(-bc+bd+3k^2+1)\\
a_0&=&ab
\end{eqnarray*}

\noindent are the coefficients of the characteristic polynomial $p(\lambda)=\lambda^4+a_3\lambda^3+a_2\lambda^2+a_1\lambda+a_0$.

Since the post-bifurcation dynamics following regular Hopf bifurcation in this modified Chen system has been recently studied in \cite{ref12}, we are interested into double Hopf bifurcations. At the double Hopf bifurcation the four eigenvalues of the linearized system must be two pairs of purely imaginary complex conjugates $\lambda_{1,2}=\pm i \sigma_1, \lambda_{3,4}=\pm i \sigma_2$ and the characteristic equation must have the form $(\lambda^2+\sigma_1^2)(\lambda^2+\sigma_2^2)=0$, which leads to the following conditions for a double Hopf bifurcation to occur:

\begin{equation}\label{bif_par}
\begin{array}{lllll}
a_3=0,\\
a_2=\sigma_1^2+\sigma_2^2>0,\\
a_1=0,\\
a_0=\sigma_1^2\sigma_2^2>0,\\
a_2^2-4a_0>0.
\end{array}
\qquad\Rightarrow\qquad
\begin{array}{llll}
c=c^*=a+b,\\
d=d^*=a+b-\displaystyle\frac{1}{b}(3k^2+1),\\
k^2>\displaystyle\frac{a+b^3+2b\sqrt{ab}}{b-2a},\\
\sigma_1,\sigma_2\in \mathbb{R},
\end{array}
\end{equation}

\noindent both for $a,b>0$ and $b>2a$ or $a,b<0$ and $b<2a$.

\section{Normal form and general post-bifurcation dynamics}

To compute the normal form of the modified Chen system \eqref{os0} we will use a perturbation technique based on the multiple scales method \cite{NB,Yu1,Yu4}.

Near the double Hopf bifurcation different time scales $T_j=\varepsilon^j t,\,j=0,1,2,\dots$ can be distinguished and
therefore the time derivative decouples as follows:

\begin{equation}\label{time}
\frac{d}{dt}=\frac{\partial}{\partial T_0}+\varepsilon \frac{\partial}{\partial T_1}+\varepsilon^2 \frac{\partial}{\partial T_2}+\varepsilon^3 \frac{\partial}{\partial T_3}\dots
\end{equation}

\noindent  Let us write the solutions of the original system \eqref{os0} and the bifurcation parameters as nonlinear expansions in $\varepsilon$:

\begin{eqnarray}\nonumber
x=\varepsilon x_1+\varepsilon^2 x_2+\varepsilon^3 x_3+O(\varepsilon^4),&\quad& y=\varepsilon y_1+\varepsilon^2 y_2+\varepsilon^3 y_3+O(\varepsilon^4),\\\label{expa}
z=\varepsilon z_1+\varepsilon^2 z_2+\varepsilon^3 z_3+O(\varepsilon^4),&\quad& w=\varepsilon w_1+\varepsilon^2 w_2+\varepsilon^3 w_3+O(\varepsilon^4),\\\nonumber
c=c^*+\varepsilon c_1+\varepsilon^2 c_2+\varepsilon^3 c_3+O(\varepsilon^4),&\quad&  d=d^*+\varepsilon d_1+\varepsilon^2 d_2+\varepsilon^3 d_3+O(\varepsilon^4),
\end{eqnarray}

\noindent where $c^*, d^*$ are the bifurcation values at the double Hopf singularity as calculated in \eqref{bif_par}. All the expansion coefficients $x_i, y_i, z_i$ and $w_i,\, i=1,2,3,\dots$ depend on the time scales $T_j,\,j=0,1,2,\dots$.

Substituting all the above expansions \eqref{time}-\eqref{expa} into the system \eqref{os0}
and collecting the terms at each order in $\varepsilon$, one gets
a sequence of differential systems for the $x_i, y_i, z_i$ and $w_i,\, i=1,2,3,\dots$:

\begin{eqnarray}\label{linw1}
O(\varepsilon)&:& \hskip 2.3cm\mathcal{L}w_1=0,\\
O(\varepsilon^2)&:& \hskip 2.3cm\mathcal{L}w_2=\Gamma_2,\label{linw2}\\
O(\varepsilon^3)&:&\hskip 2.3cm\mathcal{L}w_3=\Gamma_3,\label{linw3}
\end{eqnarray}

\noindent where $\mathcal{L}$ is the following linear operator:

\begin{eqnarray}\nonumber
\mathcal{L}&=&\frac{1}{k}\left(\frac{1}{a}\frac{\partial^4}{\partial T_0^4}+
\left(1+\frac{b-c^*}{a}\right)\frac{\partial^3}{\partial T_0^3}+\left(
\frac{k^2-bc^*}{a}+\frac{k^2}{b}+b-c^*+d^*\right)\frac{\partial^2}{\partial T_0^2}\right.\\\label{linop}
&\ &+\left.(1-bc^*+bd^*+3k^2)\frac{\partial}{\partial T_0}+b\right)
\end{eqnarray}

\noindent and the source terms $\Gamma_2$ and $\Gamma_3$  contains the nonlinear terms. The equations for $x_i, y_i$ and $z_i$ can be easily obtained only in terms of $w_i$, but here we skip all the calculation details.

The solution of the linear homogeneous problem \eqref{linw1} is given by:

\begin{equation}\label{w1sol}
w_1=\sum_{j=1}^2\left(\alpha_j(T_k)e^{i\sigma_jT_0}+\bar{\alpha}_j(T_k)e^{-i\sigma_jT_0}\right),
\end{equation}

\noindent where the fields $\alpha_j$, depending on the time scales $T_k,\,k=1,2,\dots$, lie on the center manifolds and $\bar{\alpha}_j$  represent their complex conjugate fields.
The real numbers $\sigma_j$ are the imaginary parts of the pairs of purely imaginary complex eigenvalues  calculated at the double Hopf singularity.
Suppressing the secular terms appearing in $\Gamma_2$ only by imposing $T_1=0$ and $c_1, d_1=0$ one gets the solutions $w_2$.

The solvability condition at $O(\varepsilon^3)$ leads to
the following two coupled equations for the fields $\alpha_1$ and $\alpha_2$:

\begin{equation}\label{comp_eq}
\frac{\partial \alpha_j}{\partial T_2}=L^{(j)}\alpha_j+M^{(j)}\alpha_j^2\bar{\alpha}_j+N^{(j)}\alpha_1\alpha_2\bar{\alpha}_j,
\end{equation}

\noindent where the coefficients $L^{(j)}$ are linearly dependent on the second order deviation $c_2$ and $d_2$ from the bifurcation values.

Using the polar coordinates $\alpha_j=\rho_j e^{i\vartheta_j},\ \bar{\alpha}_j=\rho_j e^{-i\vartheta_j}$,
with $\rho_j$ and $\vartheta_j$ depending on $T_2$ and separating the real and the imaginary parts, one obtains the following normal form:

\begin{eqnarray}\label{normal1}
\frac{\partial \rho_j}{\partial T_2}&=&\rho_j\left(L_{11}^{(j)}c_2+L_{12}^{(j)}d_2+M_1^{(j)}\rho_j^2+N_1^{(j)}\rho_{(-j+3)}^2\right),\\\label{normal2}
\frac{\partial \vartheta_j}{\partial T_2}&=&L_{21}^{(j)}c_2+L_{22}^{(j)}d_2+M_2^{(j)}\rho_j^2+N_2^{(j)}\rho_{(-j+3)}^2.
\end{eqnarray}

\noindent The explicit expression of the coefficients $L_{1l}^{(j)}, L_{2l}^{(j)}, M_l^{(j)}$ and $N_l^{(j)},\, l=1,2$ in terms of the parameters of the original system \eqref{os0} can be found in \cite{GCCsub11}.

The obtained normal form will be an analytical approximation of the periodic orbits at the double Hopf singularity.

The analysis of this normal form closely parallels earlier work by Yu and co-workers, in particular \cite{Yu5,Yu6}.

The stationary solutions of the equations \eqref{normal1}-\eqref{normal2} are the following:

\begin{eqnarray}
O&:&\rho_1=\rho_2=0,\label{ies}\\
H_1&:&\rho^2_1=-\displaystyle\frac{1}{M_1^{(1)}}\left(L_{11}^{(1)}c_2+L_{12}^{(1)}d_2\right),\ \rho_2=0,\label{hb1}\\
\nonumber \ &\ & \vartheta_1=L_{21}^{(1)}c_2+L_{22}^{(1)}d_2+M_2^{(1)}\rho_1^2,\\
H_2&:&\rho_1=0,\  \rho_2^2=-\displaystyle\frac{1}{M_1^{(2)}}\left(L_{11}^{(2)}c_2+L_{12}^{(2)}d_2\right),\label{hb2}\\
\nonumber \ &\ & \vartheta_2=L_{21}^{(2)}c_2+L_{22}^{(2)}d_2+M_2^{(2)}\rho_2^2,\\
T&:&\rho_1^2=\displaystyle\frac{-M_1^{(2)}\left(L_{11}^{(1)}c_2+L_{12}^{(1)}d_2\right)+N_1^{(1)}\left(L_{11}^{(2)}c_2+L_{12}^{(2)}d_2\right)}
{M_1^{(1)}M_1^{(2)}-N_1^{(1)}N_1^{(2)}},\label{qpsx}\\\label{qpsy}
\ &\ &\rho_2^2=\displaystyle\frac{-M_1^{(1)}\left(L_{11}^{(2)}c_2+L_{12}^{(2)}d_2\right)+N_1^{(2)}\left(L_{11}^{(1)}c_2+L_{12}^{(1)}d_2\right)}
{M_1^{(1)}M_1^{(2)}-N_1^{(1)}N_1^{(2)}},\\
\nonumber \ &\ & \vartheta_1=L_{21}^{(1)}c_2+L_{22}^{(1)}d_2+M_2^{(1)}\rho_1^2+N_2^{(1)}\rho_2^2,\\
\nonumber \ &\ & \vartheta_2=L_{21}^{(2)}c_2+L_{22}^{(2)}d_2+M_2^{(2)}\rho_2^2+N_2^{(2)}\rho_1^2.
\end{eqnarray}

\noindent Performing a standard linear analysis of the system \eqref{normal1}-\eqref{normal2}, it follows that the initial equilibrium state corresponding to $O$, is stable if:

\begin{equation}\label{0stable}
L_{11}^{(j)}c_2+L_{12}^{(j)}d_2<0,\ {\rm for\ } j=1,2.
\end{equation}

\noindent The conditions for the existence and stability of the stationary point $H_1$ (and consequently of a periodic solution for the system \eqref{os0}) are:

\begin{eqnarray}\label{H1stable1}
&&M_1^{(1)}<0,\\\label{H1stable2}
&&L_{11}^{(1)}c_2+L_{12}^{(1)}d_2>0,\\\label{H1stable3}
&&L_{11}^{(2)}c_2+L_{12}^{(2)}d_2-\displaystyle\frac{N_1^{(2)}}{M_1^{(1)}}\left(L_{11}^{(1)}c_2+L_{12}^{(1)}d_2\right)<0.
\end{eqnarray}

\noindent Therefore along the critical line $l_1$:

\begin{equation}\label{line1}
l_1:\qquad L_{11}^{(1)}c_2+L_{12}^{(1)}d_2=0,\qquad L_{11}^{(2)}c_2+L_{12}^{(2)}d_2 < 0
\end{equation}

\noindent the initial equilibrium bifurcates into a family of limit cycle (approximated by the periodic solution $H_1$) as follows from \eqref{0stable} and \eqref{H1stable2}.
Analogously, $H_2$ exists stable when:

\begin{eqnarray}\label{H2stable1}
&&M_1^{(2)}<0,\\\label{H2stable2}
&&L_{11}^{(2)}c_2+L_{12}^{(2)}d_2>0,\\\label{H2stable3}
&&L_{11}^{(1)}c_2+L_{12}^{(1)}d_2-\displaystyle\frac{N_1^{(1)}}{M_1^{(2)}}\left(L_{11}^{(2)}c_2+L_{12}^{(2)}d_2\right)<0,
\end{eqnarray}

\noindent and along the critical line $l_2$:

\begin{equation}\label{line2}
l_2:\qquad L_{11}^{(2)}c_2+L_{12}^{(2)}d_2=0,\qquad L_{11}^{(1)}c_2+L_{12}^{(1)}d_2 < 0
\end{equation}

\noindent the initial equilibrium bifurcates into an other family of limit cycles given by the periodic solution $H_2$.

Finally, the conditions for the existence and stability of the point $T$ are the following:

\begin{eqnarray}
&&N_1^{(1)}\left(L_{11}^{(2)}c_2+L_{12}^{(2)}d_2\right)-M_1^{(2)}\left(L_{11}^{(1)}c_2+L_{12}^{(1)}d_2\right)>0\label{Tstable1}\\
&&N_1^{(2)}\left(L_{11}^{(1)}c_2+L_{12}^{(1)}d_2\right)-M_1^{(1)}\left(L_{11}^{(2)}c_2+L_{12}^{(2)}d_2\right)>0\label{Tstable2}\\
&&M_1^{(1)}M_1^{(2)}-N_1^{(1)}N_1^{(2)}>0\label{Tstable3}\\
&&M_1^{(2)}\left(L_{11}^{(1)}c_2+L_{12}^{(1)}d_2\right)\left(N_1^{(2)}-M_1^{(1)}\right)\nonumber\\&&\hskip3cm+
M_1^{(1)}\left(L_{11}^{(2)}c_2+L_{12}^{(2)}d_2\right)\left(N_1^{(1)}-M_1^{(2)}\right)<0.\label{Tstable4}
\end{eqnarray}

\noindent From the conditions \eqref{H1stable1}-\eqref{H1stable3} and \eqref{Tstable2} it follows that along the line $l_3$:

\begin{equation}\label{l3}
l_3:\left(L_{11}^{(2)}-\frac{N_1^{(2)}}{M_1^{(1)}}L_{11}^{(1)}\right)c_2+\left(L_{12}^{(2)}-\frac{N_1^{(2)}}{M_1^{(1)}}L_{12}^{(1)}\right) d_2=0,\quad {\rm and\ \eqref{H1stable1}-\eqref{H1stable2}\ hold}
\end{equation}

\noindent the periodic solution $H_1$ bifurcates with frequency $\vartheta_2$ into a quasi-periodic solution (with a secondary Hopf bifurcation) and a 2-D torus arises.
Analogously, taking into account \eqref{H2stable1}-\eqref{H2stable3} and \eqref{Tstable1}, one individuates the line $l_4$:

\begin{equation}\label{l4}
l_4: \left(L_{11}^{(1)}-\frac{N_1^{(1)}}{M_1^{(2)}}L_{11}^{(2)}\right)c_2+\left(L_{12}^{(1)}-\frac{N_1^{(1)}}{M_1^{(2)}}L_{12}^{(2)}\right) d_2=0,\quad
{\rm and\ \eqref{H2stable1}-\eqref{H2stable2}\ hold}
\end{equation}

\noindent where the other periodic solution corresponding to $H_2$ bifurcates to the 2D motion with frequency $\vartheta_1$.

Finally, a 3D dynamical behavior can be even predicted when the following line $l_5$:

\begin{eqnarray}\label{l5}
l_5:&& \left[L_{11}^{(1)}M_1^{(2)}\left(N_1^{(2)}-M_1^{(1)}\right)+L_{11}^{(2)}M_1^{(1)}\left(N_1^{(1)}-M_1^{(2)}\right)\right]c_2+\\
&&\left[L_{12}^{(1)}M_1^{(2)}\left(N_1^{(2)}-M_1^{(1)}\right)+L_{12}^{(2)}M_1^{(1)}\left(N_1^{(1)}-M_1^{(2)}\right)\right]d_2=0\nonumber
\end{eqnarray}

\noindent is located between the lines $l_3$ and $l_4$ as follows from conditions \eqref{Tstable1}-\eqref{Tstable3}.

In the post-double-Hopf bifurcation regime, the solutions exist on a torus in phase-space.
The possible routes to chaos would most likely one of the quasiperiodic routes, i.e., either
torus doubling via period doubling of the torus, or gradual torus breakdown, or the Ruelle-Takens route
into chaos \cite{NB}. If the parameters are such that the system is locally very strongly dissipative or volume-contracting, intermittency is a possibility following bifurcations of the torus since the repulsion from the
bifurcated (and now-unstable) torus may combine with volume contraction to set up a repulsion-reinjection
intermittency scenario. However, this is less likely for most parameter sets than the previous three
quasiperiodic routes. In exceptional cases, crises may also be possible, but again,
this is much less likely for most system parameters than the quasiperiodic routes.

The modified Chen system \eqref{os0} is next integrated for various parameters sets, showing the following three main behaviors:
\begin{enumerate}
\item \noindent the trajectories fly off to infinity;
\item \noindent the trajectories evolve towards a limit cycle;
\item \noindent the trajectories evolve to a strange attractor.
\end{enumerate}

\begin{figure}
\begin{center}
\includegraphics[scale=.35]{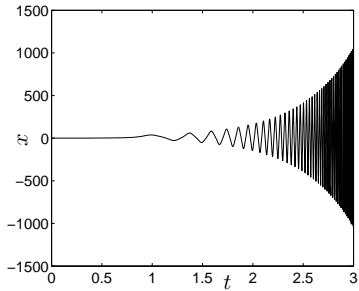}
\end{center}
\caption{\label{infi}The time behavior of the state $x$ for the system \eqref{os0}. }
\end{figure}

In all the numerical simulations the initial conditions are chosen to be close to the fixed point.
The case 1., shown in Fig. \ref{infi}, is realized at the parameter values $a=1, b=7$ and $k=9$ for which the double Hopf conditions are satisfied at
$c^*=8$ and $d^*=-35$.
Choosing $c_2=d_2=0.1$ the normal form \eqref{normal1}-\eqref{normal2} admits no stable fixed point (in particular the points $H_2$ and $T$ exist unstable, instead $H_1$ is complex) and it is not dissipative, therefore the trajectories fly off to infinity.

\begin{figure}[h]
\begin{center}
\subfigure[] {\epsfxsize=2.1 in \epsfbox{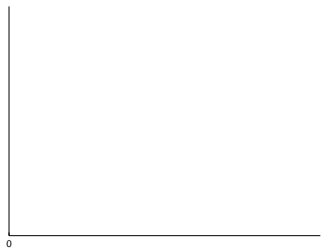}}
\end{center}
\caption{\label{perio} The trajectories evolve towards a limit cycle. }
\end{figure}

\noindent The case 2. occurs for the parameter choice $a=1, b=3, k=-6.5$ (the critical parameter values are
$c^*=4$ and $d^*=-38.5833$). Choosing the second order deviation $c_2=-19$ and $d_2=-20$ the system is strongly dissipative and
the equilibrium point $H_1$ exists stable. The other two equilibria $H_2$ and $T$ are complex. Therefore our analysis predicts that the system states evolve towards a periodic orbit, in agreement with the simulation shown in Fig. \ref{perio}.

\begin{figure}[h]
\begin{center}
\subfigure[] {\epsfxsize=2.1 in \epsfbox{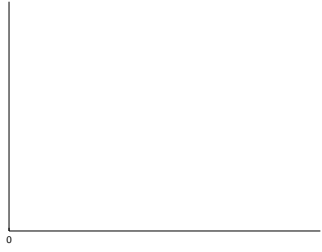}} 
\end{center}
\caption{\label{strange}  The strange attractor in the $xyz$-space. }
\end{figure}

Finally, for the choice $a=1, b=3,
k=-7$ and the second order deviations $c_2=-3$ and $d_2=-3.6667$, no stable equilibria exist for the normal form \eqref{normal1}-\eqref{normal2} (in particular $H_1$ exists unstable and the equilibrium points $H_2$ and $T$ are complex). Due to the strong dissipativity, the numerical simulation  in Fig. \ref{strange} shows that the trajectories evolve to a strange attractor.

Note that other possible route to chaos can be investigated, e.g. in \cite{GCCsub11} the post-bifurcation dynamics in the context of two intermittent routes to chaos (routes following either subcritical or supercritical Hopf or double Hopf bifurcation) are observed.

\section{Adaptive synchronization of the hyperchaotic system}

The synchronization of two chaotic/hyperchaotic systems consists into designing a controller or forcing in such a way that the motion of each system can be adjusted to a common dynamics. The intrinsic nature of the chaotic systems does not obey to this idea of synchronization due to the sensitivity to initial conditions; in fact the trajectories of two identical chaotic systems evolve to completely different dynamical behavior when starting from different initial conditions (however they are close). Nevertheless, in the last decades, various synchronization methods have been proposed which show how it is possible to synchronize this kind of systems \cite{ASL09,WGW08,GLS2,Park05,ref7,ref8}.

In our case we want to realize a complete synchronization between two identical
modified Chen system. The first one is the following master or driver system:

\begin{equation}\left\{
\begin{array}{l}\label{master}
\dot{x}_m=a(y_m-x_m)\\
\dot{y}_m=-dx_m-xz_m+cy_m-w_m-\displaystyle\frac{k^2}{b}x_m+kz_m\\
\dot{z}_m=x_my_m-bz_m-k(x_m+y_m)\\
\dot{w}_m=x_m
\end{array}\right.
\end{equation}

\noindent and the second one is the slave or response system:

\begin{equation}\left\{
\begin{array}{l}\label{slave}
\dot{x}_s=a_1(y_s-x_s)+u_1\\
\dot{y}_s=-d_1x_s-xz_s+c_1y_s-w_s-\displaystyle\frac{k_1^2}{b_1}x_s+k_1z_s+u_2\\
\dot{z}_s=x_sy_s-b_1z_s-k_1(x_s+y_s)+u_3\\
\dot{w}_s=x_s+u_4
\end{array}\right.
\end{equation}

\noindent whose evolution is guided by the controllers $u_1, u_2, u_3$ and $u_4$ and the parameters $a_1, b_1, c_1, d_1$ and $k_1$ need to be estimated in such a way that the two systems \eqref{master} and \eqref{slave} can be synchronized.

Our goal is to design suitable control functions $u_i$ and to find proper update rules for the parameters $a_1, b_1, c_1, d_1$ and $k_1$ such that the response system globally synchronizes the driver system. This is equivalent to require that the error dynamical system, obtained by
subtracting equations \eqref{master} from  \eqref{slave}:

\begin{equation}\left\{
\begin{array}{l}\label{error}
\dot{e}_1=a_1(y_s-x_s)-a(y_m-x_m)+u_1\\
\dot{e}_2=-d_1x_s+dx_m-xz_s+x_mz_m+c_1y_s-cy_m-w_s+w_m\\
\qquad-\displaystyle\frac{k_1^2}{b_1}x_s+\displaystyle\frac{k^2}{b}x_m+k_1z_s-kz_m+u_2\\
\dot{e}_3=x_sy_s-x_my_m-b_1z_s+bz_m-k_1(x_s+y_s)+k(x_m+y_m)+u_3\\
\dot{e}_4=x_s-x_m+u_4
\end{array}\right.
\end{equation}

\noindent where $e_1=x_s-x_m, e_2=y_s-y_m, e_3=z_s-z_m, e_4=w_s-w_m$,
is asymptotically stable to the origin, i.e. $\lim_{t\rightarrow \infty}||e(t)||=0$,
 $e=[e_1\ e_2\ e_3\ e_4]^T$.

\vskip1cm
\textbf{Theorem}: For any initial conditions, the two systems \eqref{master} and \eqref{slave} are globally asymptotically synchronized by the following control law:

\begin{equation}
\begin{array}{l}\label{control}
u_1=-(h_1-a_1)e_1-(a_1-d_1-z_s)e_2-e_4\\
u_2=-(h_2+c_1)e_2-e_1e_3-k_1\left(\displaystyle\frac{k}{b}x_m-\displaystyle\frac{k_1}{b_1}x_s\right)\\
u_3=-(h_3-b_1)e_3+k_1e_1-y_me_1\\
u_4=-h_4e_4+e_2
\end{array}
\end{equation}

\noindent where $h_i$ are positive scalars (called control gains) and by the parameter update rules:

\begin{equation}\left\{
\begin{array}{l}\label{parameterRule}
\dot{e}_a=-(y_m-x_m)e_1\\
\dot{e}_b=z_me_3\\
\dot{e}_c=-y_me_2\\
\dot{e}_d=x_m e_2\\
\dot{e}_k=\displaystyle\frac{k}{b}x_m e_2-z_me_2+(x_m+y_m)e_3
\end{array}\right.
\end{equation}

\noindent where $e_a=a_1-a, e_b=b_1-b, e_c=c_1-c, e_d=d_1-d, e_k=k_1-k$.

\vskip.5cm
\textit{Proof}:
The proof of the Theorem is based on the Lyapunov stability theory. Let us choose the following Lyapunov function:

\begin{equation}\label{Lyap}
V=\frac{1}{2}\left(e_1^2+e_2^2+e_3^2+e_4^2+e_a^2+e_b^2+e_c^2+e_d^2+e_k^2\right),
\end{equation}

\noindent which is positive definite. Calculating the time derivative of the Lyapunov function \eqref{Lyap} along the trajectories of the error system \eqref{error} one obtains:

\begin{eqnarray}\label{LyapDiff}\nonumber
\dot{V}&=&\dot{e}_1e_1+\dot{e}_2e_2+\dot{e}_3e_3+\dot{e}_4e_4+\dot{e}_ae_a+\dot{e}_be_b+\dot{e}_ce_c+\dot{e}_de_d+\dot{e}_ke_k\\\nonumber
&=&e_1\left(a_1(y_s-x_s)-a(y_m-x_m)+u_1\right)+e_2\left(-d_1x_s+dx_m-xz_s+x_mz_m\right.\\
&+&\left.c_1y_s-cy_m-w_s+w_m
-\displaystyle\frac{k_1^2}{b_1}x_s+\displaystyle\frac{k^2}{b}x_m+k_1z_s-kz_m+u_2\right)\\\nonumber
&+&
e_3\left(x_sy_s-x_my_m-b_1z_s+bz_m-k_1(x_s+y_s)+k(x_m+y_m)+u_3\right)\\\nonumber&+&
e_4\left(x_s-x_m+u_4\right)+\dot{e}_a(a_1-a)+\dot{e}_b(b_1-b)+\dot{e}_c(c_1-c)+\dot{e}_d(d_1-d)\\&+&\dot{e}_k(k_1-k).\nonumber
\end{eqnarray}

\noindent Substituting equations \eqref{control} and \eqref{parameterRule} into the expression \eqref{LyapDiff} for $\dot{V}$ and noting that the following equalities hold:

\begin{eqnarray}\label{position}
-x_sz_s+x_mz_m&=&-z_se_1-x_me_3,\\\nonumber
x_sy_s-x_my_m&=&x_se_2+y_me_1,
\end{eqnarray}

\noindent  one gets:

\begin{equation}\label{LyapDiffSimpl}
\dot{V}=-\sum_{i=1}^4 h_i e_i^2=-e^TPe, \qquad {\rm where\ } P={\rm diag}\{h_1, h_2, h_3, h_4\}.
\end{equation}

\noindent Since $\dot{V}$ is negative semidefinite $e_1, e_2, e_3, e_4, e_a, e_b, e_c, e_d, e_k \in \mathcal{L}_{\infty}$ and from the error system \eqref{error} it follows that $\dot{e}_1, \dot{e}_2, \dot{e}_3, \dot{e}_4\in \cal{L}_{\infty}$. Given $\lambda_{\rm min}(P)$ the minimum eigenvalue of the matrix $P$, one gets:

\begin{equation}
\int_0^t \lambda_{\rm min}(P)||e||^2 {\rm d}t\leq \int_0^t e^TPe{\rm d}t\leq \int_0^t -\dot{V}{\rm d}t=V(0)-V(t)\leq V(0).
\end{equation}

\noindent Therefore $e_1, e_2, e_3, e_4\in \mathcal{L}_2$ and the hypotheses of the Barbalat's lemma (see \cite{SL91} for details) are satisfied. Thus  $\lim_{t\rightarrow \infty}||e(t)||=0$ and the proof is completed. 

\hskip11cm$\square$
\vskip.2cm
To test the effectiveness of the proposed adaptive synchronization scheme, we show two numerical examples in which the modified Chen system has been chosen both in its chaotic and hyperchaotic regime.
By choosing the parameters $a=1, b=3, c=1, d=-49, k=-7$ the master system evolves to the chaotic attractor shown in Fig.\ref{strange} (we have checked that only one Lyapunov exponent is positive). Let us assume that the starting points for the master and the slave systems are respectively $(x_m(0), y_m(0),z_m(0), w_m(0))=(0.1,0.3,0.01,.2)$ and $(x_s(0), y_s(0),z_s(0), w_s(0))=(-0.1,0.3,-0.01,-0.2)$.

\begin{figure}
\begin{center}
\includegraphics[scale=.6]{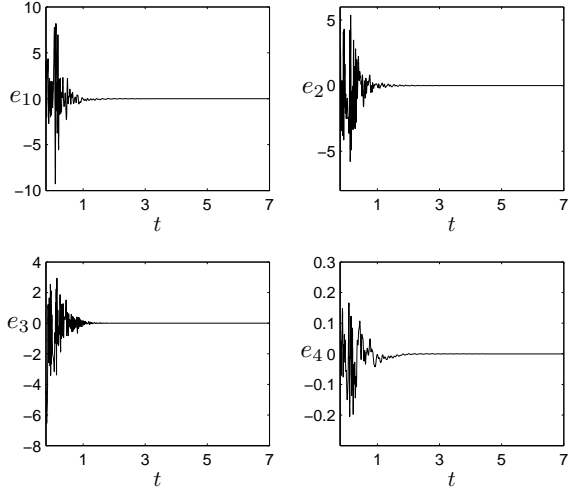}
\end{center}
\caption{\label{controlAttrac} Chaos synchronization: the trajectories of the error dynamical system asymptotically converge to the origin.}
\end{figure}

\begin{figure}
\begin{center}
\includegraphics[scale=.7]{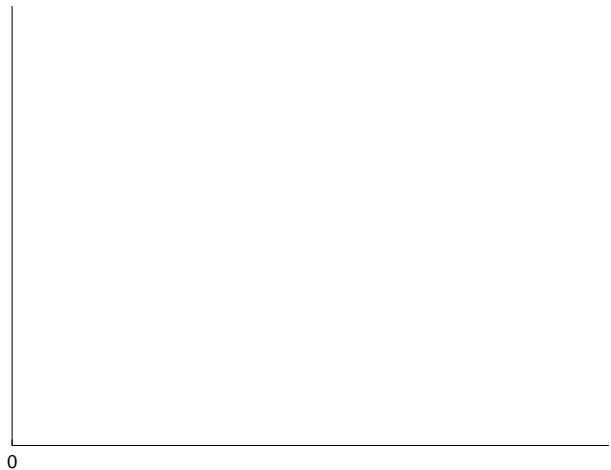}
\end{center}
\caption{\label{controlParam}The master and the slave system have been synchronized to the chaotic attractor shown in the last subfigure. The parameters of the slave system asymptotically converge to the master system parameter.}
\end{figure}

Given the initial conditions for the uncertain parameters of the response system $a_1(0)=3, b_1(0)=5, c_1(0)=2, d_1(0)=-30, k_1(0)=-5$, the adaptive control laws \eqref{control} with control gains $(h_1, h_2, h_3, h_4)=(5,7,6,5)$ realize the chaotic synchronization of the systems \eqref{master} and \eqref{slave} as shown in Figures \ref{controlAttrac}-\ref{controlParam}.

\begin{figure}
\begin{center}
\includegraphics[scale=.6]{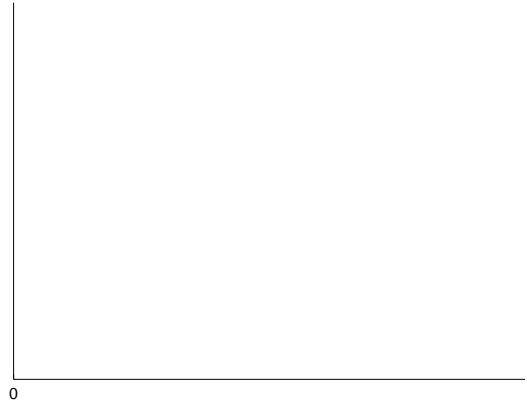}
\end{center}
\caption{\label{controlHyper}Hyperchaotic synchronization: the error dynamical system asymptotically converges to the origin.}
\end{figure}

\begin{figure}
\begin{center}
\includegraphics[scale=.7]{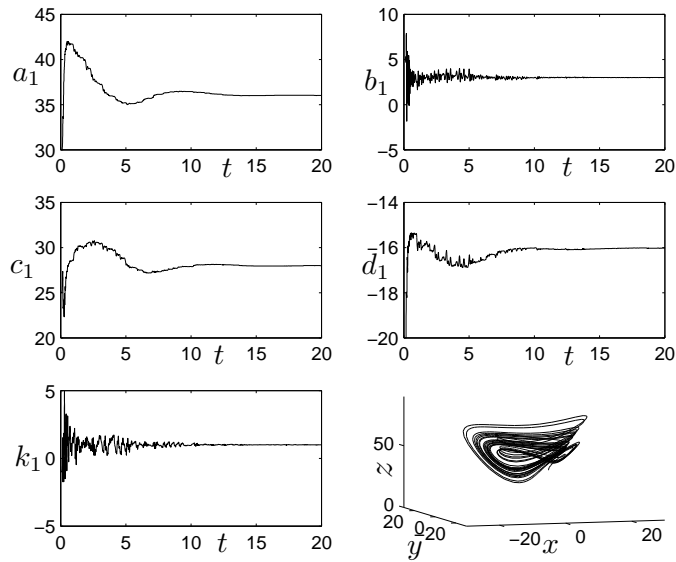}
\end{center}
\caption{\label{controlParamHyper}The estimates of the slave system parameters converge to the master system parameters. Both the systems \eqref{master} and \eqref{slave} evolve to the hyperchaotic attractor shown in the last subfigure.}
\end{figure}

In the second numerical test the master systems has been chosen in its hyperchaotic regime, with the parameters $a=36, b=3, c=28, d=-16, k=1$ (the two positive Lyapunov exponents are shown in Fig.\ref{f1}). Let us choose
the initial conditions of the master and the slave systems respectively  as $\left(x_m(0), y_m(0),\right.$ $\left.z_m(0), w_m(0)\right)=(0.3,0.03,0.1,0.2)$ and $\left(x_s(0), y_s(0),z_s(0),\right.$ $\left.w_s(0))=(-0.1,0.3, 0.01,-0.3\right)$.
Given the initial estimated parameters $a_1(0)=30, b_1(0)=5,$ $c_1(0)=25, d_1(0)=-20, k_1(0)=-1$ and the control gains $(h_1, h_2, h_3, h_4)=(5,4,6,5)$ the effectiveness of the hyperchaotic synchronization is shown in Figures \ref{controlHyper}-\ref{controlParamHyper}.

\section{Conclusions}
In order to understand the onset of hyperchaotic behavior recently
observed in many dynamical systems, we have constructed and analyzed the generalized double Hopf normal form in the modified Chen system which reveals possible regimes of
periodic solutions, two-period tori, and three-period tori in parameter space.
Numerical simulations are provided showing agreement with the predictions from
the normal form.
In a future work other possible routes to chaos will be investigated, as further bifurcations of the post-supercritical-Hopf two- and three-tori via either torus doubling or breakdown.
Moreover, an adaptive synchronization of the chaotic/hyperchaotic system trajectories has been globally realized and the effectiveness of this control strategy has been numerically illustrated.

\end{document}